\documentclass[aps,prl,preprint,superscriptaddress,showpacs,preprintnumbers]{revtex4-1}
\usepackage[colorlinks=true, pdfstartview=FitV, linkcolor=red, citecolor=blue, urlcolor=black, pdftitle={},pdfauthor={Tomoya Hayata},pdfsubject={}, pdfkeywords={}]{hyperref}
\usepackage[pdftex]{graphicx}
\usepackage{setspace,bm}
\usepackage{latexsym,amssymb,amsmath,mathrsfs}
\usepackage{color}
\graphicspath{{./figure/}}

\newcommand{\cL}{{\cal L}}
\newcommand{\cA}{{\cal A}}

\newcommand{\p}{\partial}
\newcommand{\ve}{\varepsilon}

\newcommand{\be}{\begin{equation}}      
\newcommand{\ee}{\end{equation}}      
\newcommand{\bea}{\begin{eqnarray}}      
\newcommand{\eea}{\end{eqnarray}}

\newcommand{\tr}{\mathrm{tr}}
\newcommand{\im}{\mathrm{i}}
\newcommand{\pf}{\mathop{\mathrm{Pf}}}

\begin{document}
\title{Kinetic Theory and Anomalous Transports in the Presence of Nonabelian Phase-Space Berry Curvatures}

\preprint{RIKEN-QHP-280, RIKEN-STAMP-33}
\author{Tomoya Hayata}
\affiliation{
Department of Physics, Chuo University, 1-13-27 Kasuga, Bunkyo, Tokyo, 112-8551, Japan 
}
\author{Yoshimasa Hidaka}
\affiliation{
Theoretical Research Division, Nishina Center, RIKEN, Wako, Saitama 351-0198, Japan
}

\date{\today}

\begin{abstract}
We construct the kinetic theory in ($1+2d$)-dimensional phase space and time when all abelian and nonabelian phase-space Berry curvatures are nonzero.
Then we calculate anomalous transports induced by the Berry curvatures on the basis of the kinetic theory.
As an example, we study anomalous charge and spin transports induced by the SU($2$) Berry curvatures. 
We also derive the topological effective theory to reproduce the transports in insulators calculated from the kinetic theory.
Such an effective theory is given by the nonabelian phase space Chern-Simons theory.

\end{abstract}

\maketitle

{\it Introduction.} 
The wave function in quantum mechanics often acquires the non-trivial phase under an adiabatic and cyclic process, which is known as the so called Berry phase~\cite{Berry45}.
The Berry phase and the Berry curvature describe torsion of the wave function in a closed manifold composed by parameters of the Hamiltonian. 
They provide us a universal description  of anomalous transport phenomena, represented by the seminal works on 
 the quantum Hall effect~\cite{PhysRevLett.45.494,PhysRevB.23.5632,PhysRevLett.49.405,PhysRevLett.51.51,PhysRevB.31.3372,KOHMOTO1985343}, 
 and the adiabatic charge pumping~\cite{PhysRevB.27.6083,PhysRevB.47.1651,RevModPhys.66.899}. 

The effect of the Berry curvature has been incorporated into kinetic theory to study anomalous transports in metals as well as insulators~\cite{PhysRevB.59.14915,RevModPhys.82.1959}. 
An example of this is the intrinsic contribution to the anomalous Hall effect~\cite{PhysRev.95.1154,PhysRevLett.88.207208,JPSJ.71.19,PhysRevLett.93.206602,RevModPhys.82.1539}. 
Later, the modified kinetic theory has been applied to study anomalous transports in Dirac and Weyl semimetals~\cite{Liu864,PhysRevLett.113.027603,Xu613,Lu622,PhysRevX.5.031013} such as the chiral magnetic effect~\cite{NIELSEN1983389,PhysRevD.78.074033,PhysRevB.88.104412}. 
Interestingly, it has been shown that the Berry curvature of a Weyl fermion is closely related to the triangle anomaly, and the modified kinetic theory has been applied to study anomalous transports in high-energy physics under the name of the chiral kinetic theory~\cite{PhysRevLett.109.162001,PhysRevLett.109.181602,PhysRevD.87.085016,PhysRevLett.110.262301}.

The topological nature of the anomalous transports induced by the Berry curvatures can be understood from the relation with topological field theory.
For example, the low-energy effective theory of the quantum Hall state is given by the ($1+2$)-dimensional Chern-Simons theory~\cite{PhysRevLett.62.82,PhysRevB.44.5246,ZHANG}, 
and that of topological insulators~\cite{Bernevig1757,Konig766,Hsieh2008,Xia2009,Zhang2009,Chen2009,RevModPhys.82.3045} is given by the so called $\theta$ term in $1+3$ dimensions~\cite{PhysRevB.78.195424,PhysRevLett.102.146805,RevModPhys.83.1057}.
Those topological effective theory in real space are generalized to phase space for describing the electromagnetic responses induced by the phase-space Berry curvatures~\cite{PhysRevB.78.195424,PhysRevX.5.021018}.
Recently in Ref.~\cite{Hayata:2016wgy}, we have shown, for the abelian Berry curvatures, that such a topological effective theory is completely equivalent to the kinetic theory, and can reproduce the anomalous transports obtained from the modified kinetic theory~\cite{PhysRevLett.102.087602}.

In contrast to the abelian case, the anomalous transports in the presence of nonabelian phase space Berry curvatures await full understanding,
which is necessary to study the nonabelian generalization of the anomalous transports such as the adiabatic charge pumping, the chiral magnetic effect, and the anomalous Hall effect.
There have been attempts to derive a kinetic theory in a multi-band case with the nonabelian Berry curvatures from the derivative expansion of the Wigner function~\cite{PhysRevB.72.085110,Shindou2005399,PhysRevB.77.035110,0953-8984-20-19-193202}.
Such a kinetic theory has been applied to study the intrinsic contribution to the spin Hall effect~\cite{Murakami1348,PhysRevLett.93.156804}. 
However, neither the full expression of the anomalous transports nor the relation to the Chern-Simons theory has not been derived unlike the case of the abelian Berry curvature.

In this Letter, we construct the kinetic theory when all abelian and nonabelian phase-space Berry curvatures in $1+2d$ dimensions are nonzero.
For this purpose we generalize the kinetic theory in the presence of the nonabelian gauge fields, which has been developed in the context of quark-gluon-plasma to study the nonequilibrium dynamics of the plasma such as thermalization with color exchange interactions~\cite{Wong1970,PhysRevLett.51.351,HEINZ198548,HEINZ1986148,ELZE198981,PhysRevLett.72.3461,PhysRevD.50.4209,Blaizot:2001nr,Litim:2001db}.
We generalize it to incorporate the interactions with the nonabelian gauge fields (Berry curvatures) in momentum space and study transports induced by the Berry curvatures.
As an example, we elaborate the anomalous transports induced by the SU($2$) Berry curvatures. 
We derive the nonabelian version of the adiabatic charge pumping, the chiral magnetic effect and the anomalous Hall effect, and their analogue in the spin transports.
Finally, we construct an effective theory to reproduce the transports obtained from the kinetic theory.
We show that such an effective theory is the nonabelian generalization of the phase space Chern-Simons theory~\cite{PhysRevX.5.021018,Hayata:2016wgy}.

{\it Kinetic theory and non-abelian Berry connections.}
We consider the semiclassical dynamics in ($1+2d$)-dimensional phase space and time.
The action has the form:
\be
\begin{split}
{S} =\int dt \Bigl( &\dot{\bm x}\cdot(\bm p+\bm A)+\dot{\bm p}\cdot\bm a
\\
+&\dot{\bm x}\cdot q^{\hat{a}}\bm A^{\hat{a}}+\dot{\bm p}\cdot q^{\hat{a}}\bm a^{\hat{a}}-\ve+A_t+q^{\hat{a}}A_{~t}^{\hat{a}}\Bigr),
\end{split}
\label{eq:action}
\ee
where $\ve$ is an energy eigenvalue of $N$-fold degenerate Bloch states $|u^{i}(t,\xi_a,q^{\hat{a}})\rangle$ ($i=1,\ldots, N$), which depend on time $t$ and phase space coordinates $\xi_a=(\xi_1,\ldots,\xi_{2d})=(\bm x, \bm p)$ with $\bm x=(x_1,\ldots,x_d)$ and $\bm p=(p_1,\ldots,p_d)$, and also depend on ``color" charges $q^{\hat{a}}$ ($\hat{a}=1,\ldots, N^2-1$). 
We here introduce the ``color" charges as the dynamical degrees of freedom to describe the dynamics of the ``color" currents $j^{\hat{a}}_{~b}$, whose expression in kinetic theory is given later.
To keep the gauge invariance of the action~\eqref{eq:action}, $q^{\hat{a}}$ satisfies
\be
 \dot{q}^{\hat{a}}=-f^{\hat{a}\hat{b}\hat{c}}\cA_{~t}^{\hat{b}}q^{\hat{c}}-f^{\hat{a}\hat{b}\hat{c}}\dot{\xi}_j\cA_{~j}^{\hat{b}}q^{\hat{c}} .
\ee
$A_t$, $\bm A=(A_1,\ldots,A_d)$, and $\bm a=(a_1,\ldots,a_d)$ are the abelian Berry connections and defined as
$A_t=\tr\langle u^i| \im\p_t|u^j\rangle/N$, $\bm A=\tr\langle u^i|\im\nabla_{\bm x}|u^j\rangle/N$, and $\bm a=\tr\langle u^i|\im\nabla_{\bm p}|u^j\rangle/N$. 
In the same manner, 
$A_{~t}^{\hat{a}}$, $\bm A^{\hat{a}}=(A_{~1}^{\hat{a}},\ldots,A_{~d}^{\hat{a}})$, and $\bm a^{\hat{a}}=(a_{~1}^{\hat{a}},\ldots,a_{~d}^{\hat{a}})$ are the nonabelian Berry connections and defined as
$A_{~t}^{\hat{a}}=2\tr\left[ t^{\hat{a}}\langle u^i| \im\p_t|u^j\rangle\right]$, 
$\bm A^{\hat{a}}=2\tr\left[ t^{\hat{a}}\langle u^i|\im\nabla_{\bm x}|u^j\rangle\right]$, and 
$\bm a^{\hat{a}}=2\tr\left[ t^{\hat{a}}\langle u^i|\im\nabla_{\bm p}|u^j\rangle\right]$, with $t^{\hat{a}}$ ($\hat{a}=1,\ldots, N^2-1$) being the generators of the Lie algebras of SU($N$) group normalized as $\tr\, t^{\hat{a}}t^{\hat{b}}=\delta^{\hat{a}\hat{b}}/2$. 
$f^{\hat{a}\hat{b}\hat{c}}$ are the structure constants of the Lie algebras, whose generators are $t^{\hat{a}}$ ($[t^{\hat{a}},t^{\hat{b}}]=i f^{\hat{a}\hat{b}\hat{c}}t^{\hat{c}}$).
In the presence of the external electromagnetic fields, their gauge potentials are also introduced to the action~\eqref{eq:action}, which can be absorbed into $A_t$ and $\bm A$.
In the following, we do not distinguish the electromagnetic gauge potentials (field strengths) and the real-space abelian Berry connections (curvatures) unless otherwise stated. 
We employ the Einstein convention for repeated indices. 
We define the phase space and time coordinates and the generalized connections as  $\xi_\mu= (\xi_0,\xi_a) = (t,\xi_a)$,
$q^{\hat{\nu}}=(q^{\hat{0}},q^{\hat{b}})=(1,q^{\hat{b}})$, 
and ${\cal A}_{~\mu}^{\hat{\nu}}=({\cal A}_{~0}^{\hat{\nu}},{\cal A}_{~a}^{\hat{\nu}})=({\cal A}_{~t}^{\hat{\nu}},{\cal A}_{~a}^{\hat{\nu}})$ ($\hat{\nu}=0,1,\ldots,N^2-1$) with $\xi_a=(\bm{x},\bm{p})$, ${\cal A}_{~t}^{\hat{0}}=-\ve+A_t$, ${\cal A}_{~a}^{\hat{0}}= (\bm{p}+\bm{A},\bm{a})$, ${\cal A}_{~t}^{\hat{b}}=A_{~t}^{\hat{b}}$, and ${\cal A}_{~a}^{\hat{b}}=(\bm{A}^{\hat{b}},\bm{a}^{\hat{b}})$. 
The unhatted Greek indices such as $\mu$, and $\nu$ run from $0$ to $2d$.
The unhatted Roman indices such as $a$, and $b$ run from $1$ to $2d$.
The hatted Greek indices such as $\hat{\mu}$, and $\hat{\nu}$ run from $0$ to $N^2-1$.
The hatted Roman indices such as $\hat{a}$, and $\hat{b}$ run from $1$ to $N^2-1$.
Following Refs.~\cite{PhysRevX.5.021018,Hayata:2016wgy}, we absorb the energy $\ve$ and momentum $\bm p$ into the abelian Berry connections. 
Using these variables, we can write the action as the topological form:
$S = \int q^{\hat{\nu}}{\cal A}_{~\mu}^{\hat{\nu}} d\xi_\mu$,
where $q^{\hat{0}}=1$.
As will be seen below,  this topological form implies that the anomalous transports are expressed by using the $(1+2d)$-dimensional Chern-Simons theory for ${\cal A}_{~\mu}^{\hat{\nu}}$.
The equation of motion reads 
\be
q^{\hat{\mu}}\omega_{~ab}^{\hat{\mu}}\dot{\xi}_b= -q^{\hat{\nu}}\omega_{~at}^{\hat{\nu}}, 
\label{eq:ceom1}
\ee
where $\omega_{~\nu\lambda}^{\hat{\mu}}$ are the generalized Berry curvatures and defined as 
\be
\omega_{~\nu\lambda}^{\hat{\mu}}=\frac{\p {\cal A}_{~\lambda}^{\hat{\mu}}}{\p \xi_\nu}-\frac{\p {\cal A}_{~\nu}^{\hat{\mu}}}{\p \xi_\lambda}
+f^{\hat{\mu}\hat{\alpha}\hat{\beta}}{\cal A}_{~\nu}^{\hat{\alpha}}{\cal A}_{~\lambda}^{\hat{\beta}},
\label{eq:berry_curvature_nonabelian}
\ee
where $f^{\hat{\mu}\hat{\alpha}\hat{\beta}}$ are the structure constants with $f^{\hat{0}\hat{\alpha}\hat{\beta}}=f^{\hat{\alpha}\hat{0}\hat{\beta}}=f^{\hat{\alpha}\hat{\beta}\hat{0}}=0$.
It will be helpful to explicitly show the classical equation of motion in the standard notation in $d=3$. 
To avoid the lengthy expression, we here neglect the mixed-space Berry curvatures such as $\Omega^{\hat{\mu}}_{~x_ip_j}$:
\bea
 \dot{x}_i&=&\;\;\,\p_{p_i}\ve
 -\epsilon_{ijk}\dot{p}_j b_{~k}^{\hat{\mu}}q^{\hat{\mu}}-e_{~i}^{\hat{\mu}}q^{\hat{\mu}}, 
 \label{eq:eom_x}
 \\
 \dot{p}_i&=&-\p_{x_i}\ve 
 +\epsilon_{ijk}\dot{x}_j B_{~k}^{\hat{\mu}}q^{\hat{\mu}}+E_{~i}^{\hat{\mu}}q^{\hat{\mu}},
 \label{eq:eom_p}
\eea
where $E_{~i}^{\hat{\mu}}=\Omega_{~x_it}^{\hat{\mu}}$, and $e_{~i}^{\hat{\mu}}=\Omega_{~p_it}^{\hat{\mu}}$
($B_{~i}^{\hat{\mu}}=\epsilon_{ijk}\Omega_{~x_jx_k}^{\hat{\mu}}/2$, and $b_{~i}^{\hat{\mu}}=\epsilon_{ijk}\Omega_{~p_jp_k}^{\hat{\mu}}/2$) 
are the electric and emergent electric fields (magnetic and emergent magnetic fields) in real and momentum space, respectively. 
The definition of $\Omega_{~\nu\lambda}^{\hat{\mu}}$ is given by Eq.~\eqref{eq:berry_curvature_nonabelian} with replacing ${\cal A}_{~\nu}^{\hat{\mu}}$ by $A_{~\nu}^{\hat{\mu}}$ (Those with $\hat{\mu}=0$ are equal to the U(1) electromagnetic fields and the abelian Berry curvatures). 
The second and third terms in Eq.~\eqref{eq:eom_p} with $\hat{\mu}\neq0$ are the nonabelian generalization of the electromagnetic Lorentz and Coulomb forces, and 
the momentum analogue of them is introduced in Eq.~\eqref{eq:eom_x}. 
Similarly, in the presence of the mixed-space Berry curvatures, the mixed-space analogue of the Lorentz force appears, which can be introduced by generalizing the abelian case as $\Omega_{\nu\lambda}\rightarrow q^{\hat{\mu}}\Omega_{~\nu\lambda}^{\hat{\mu}}$ in Refs.~\cite{RevModPhys.82.1959,Hayata:2016wgy}.

We assume that $\det(q^{\hat{\mu}}\omega^{\hat{\mu}})$ is nonzero,
and then, the equation of motion~\eqref{eq:ceom1} is written as 
\be
\dot{\xi}_a=-\left[q^{\hat{\mu}}\omega^{\hat{\mu}}\right]^{-1}_{ab}q^{\hat{\nu}}\omega_{~bt}^{\hat{\nu}} .
\label{eq:ceom2}
\ee
Like the abelian case, the invariant volume element reads $d^dxd^dp d^{N^2-1}q\sqrt{\det(q^{\hat{\mu}}\omega^{\hat{\mu}})}/(2\pi)^d$, 
where the integration measure of the color space is chosen so as for the Casimir invariants of the Lie algebras to be constants of motion with the normalization $\int d^{N^2-1}q1=1$~\cite{Litim:2001db,Cho:2008wc}.
We generalize color space to include $q^{\hat{0}}$ ($q^{\hat{0}}$ itself is a constant of motion, namely, $q^{\hat{0}}=1$), and write the integration measure as $d^{N^2}q$.
Since $\omega$ is a skew symmetric matrix, whose determinant is written as $\det(q^{\hat{\mu}}\omega^{\hat{\mu}})=\pf(q^{\hat{\mu}}\omega^{\hat{\mu}})^2$ with the Pfaffian 
\be
\pf(q^{\hat{\mu}}\omega^{\hat{\mu}})
=\frac{1}{2^{d}d!}\epsilon_{a_1\cdots a_{2d}}q^{\hat{\mu}_1}\omega^{\hat{\mu}_1}_{~a_1a_2}\cdots q^{\hat{\mu}_d}\omega^{\hat{\mu}_d}_{~a_{2d-1}a_{2d}} ,
\ee
where $\epsilon_{a_1\cdots a_{2d}}$ is the totally anti-symmetric tensor ($\epsilon_{12\cdots2d}=1$). 
We note that the determinant of a real skew matrix is always nonnegative.
We find that the inverse matrix is given as~\cite{Hayata:2016wgy} 
\be
\left[q^{\hat{\mu}}\omega^{\hat{\mu}}\right]^{-1}_{ab}
=\frac{\epsilon_{baa_1\cdots a_{2d-2}}}{2^{d-1}(d-1)!\pf(q^{\hat{\nu}}\omega^{\hat{\nu}})}q^{\hat{\mu}_1}\omega^{\hat{\mu}_1}_{~a_1a_2}\cdots q^{\hat{\mu}_{d-1}}\omega^{\hat{\mu}_{d-1}}_{~a_{2d-3}a_{2d-2}} .
\label{eq:poisson_2d}
\ee

In our kinetic theory, the current can be calculated by averaging the velocity of particles, $\dot{\xi}_a$, 
over phase and color space $(\xi_a,q^{\hat{\mu}})$ with the invariant volume element. 
We consider the distribution function $n(t,\xi_a)$, which is independent of the color charges such as the thermal equilibrium distribution.
Then by using Eqs.~\eqref{eq:ceom2} and~\eqref{eq:poisson_2d}, 
the current density in phase space $j_{~a}^{\hat{\mu}}(t,\xi_a)$ is given as, by integrating over color space,
\bea
j_{~a}^{\hat{\mu}}(t,\xi_a) 
&=& \int \frac{d^{N^2}q}{(2\pi)^d}\sqrt{\det(q^{\hat{\nu}}\omega^{\hat{\nu}})}q^{\hat{\mu}}\dot{\xi}_a n(t,\xi)
\nonumber
\\
&=& \frac{(-1)^\nu c_{\hat{\mu}\hat{\mu}_1\ldots\hat{\mu}_d}\epsilon_{aba_1\cdots a_{2d-2}}}{(2\pi)^d2^{d-1}(d-1)!}\omega^{\hat{\mu}_1}_{~a_1a_2}\cdots\omega^{\hat{\mu}_{d-1}}_{~a_{2d-1}a_{2d-2}}\omega^{\hat{\mu}_d}_{~b t}n(t,\xi) ,
\label{eq:anomalouscurrent_2d}
\eea
where $\dot{\xi}_a$ is the solution of the equation of motion~\eqref{eq:ceom2}, and
we introduced  the sign of the Pfaffian $(-1)^\nu= \pf(q^{\hat{\nu}}\omega^{\hat{\nu}})/\sqrt{\det(q^{\hat{\nu}}\omega^{\hat{\nu}})}$ (In our kinetic regime, the sign is negative~\cite{Hayata:2016wgy}).
$c_{\hat{\mu}\hat{\mu}_1\ldots\hat{\mu}_{d}}=\int d^{N^2}q q^{\hat{\mu}}q^{\hat{\mu}_1}\cdots q^{\hat{\mu}_d}$ is a symmetric tensor and invariant under the adjoint action of the Lie group. 
Since $c_{\hat{0}\hat{\mu}_1\ldots\hat{\mu}_{m}}=c_{\hat{\mu}_1\ldots\hat{\mu}_{m}}$, whether $c_{\hat{\mu}\hat{\mu}_1\ldots\hat{\mu}_{d}}$ becomes nonzero or not is 
determined by $c_{\hat{a}_1\ldots\hat{a}_{m}}=\int d^{N^2-1}q q^{\hat{a}_1}\cdots q^{\hat{a}_m}$ ($m=1,\ldots, d+1$), which are written only by using the Casimir invariants of the Lie algebras~\cite{Cho:2008wc}.
Integrating Eq.~\eqref{eq:anomalouscurrent_2d} with respect to $p_i$, we obtain the current in real space.
Also the local charge density is given as $j_{~0}^{\hat{\mu}}=\int d^{N^2}q\sqrt{\det(q^{\hat{\nu}}\omega^{\hat{\nu}})}q^{\hat{\mu}}n(t,\xi)/(2\pi)^d$.
For a band insulator ($n(t,\xi_a) =1$), we have 
\bea
j_{~0}^{\hat{\mu}}(\xi)&=&\frac{(-1)^\nu}{2^{d}d!}c_{\hat{\mu}\hat{\mu}_1\ldots\hat{\mu}_{d}}\epsilon_{a_1\cdots a_{2d}}\omega^{\hat{\mu}_1}_{~a_1a_2}\cdots \omega^{\hat{\mu}_d}_{~a_{2d-1}a_{2d}}  .
\label{eq:polarization_2d}
\eea

{\it Anomalous charge and spin transports.}  
The general expression of anomalous transports is given by Eq.~\eqref{eq:anomalouscurrent_2d}.
We here elaborate the ones induced by the interplay of the SU($2$) phase-space Berry curvatures and the U($1$) electromagnetic fields.
The SU($2$) Berry curvatures arise e.g., in the Luttinger model~\cite{PhysRev.102.1030}: ${\cal H}=\sum_{a=1}^5 d_a \Gamma_a$, where $\Gamma_a$ are the generators of the SO($5$) Clifford algebras. The standard Berry curvatures in momentum space are obtained by the pull back from those in the five-dimensional $\bm d$ space to momentum space~\cite{0953-8984-20-19-193202,Murakami1348,PhysRevLett.93.156804}. 
The SU($2$) Berry curvatures in phase space and time are also  obtained by the pull back to mixed space when $\bm d$ depends on $\bm x$ and $t$, e.g., by applying strain spatial gradient and lasers.

First we consider the adiabatic charge pumping in the presence of time-periodic perturbation.
By integrating over the period of perturbation $T$ with the Fermi-Dirac distribution $n(\bm p)$ in the vanishing external fields, 
in $d=3$, it is given as
\bea
P_i(x_i)
&&\,=-e\int \frac{dtd^3p}{(2\pi)^3}j_{~i}^{\hat{0}}(t,\bm x,\bm p)
\nonumber
\\
&&
\begin{split}
=e\int \frac{dtd^3p}{(2\pi)^3}\Bigl[&
\frac{1}{2}\left(\delta_{ij}\Omega^{\hat{a}}_{~x_kp_k}-\Omega^{\hat{a}}_{~p_ix_j}\right)e^{\hat{a}}_{~j}
\\
-&\frac{1}{4}\epsilon_{ikl}\epsilon_{j\bar{m}\bar{n}}\Omega^{\hat{a}}_{~p_{\bar{m}}x_k}\Omega^{\hat{a}}_{~p_{\bar{n}}x_l}\p_{p_j}\ve
\\
+&\frac{1}{2}B^{\hat{a}}_{~i}\bm b^{\hat{a}}\cdot\nabla_{p}\ve
+\frac{1}{2}eB_{i} \bm e^{\hat{a}}\cdot \bm b^{\hat{a}}
\\
+&\frac{1}{2}\epsilon_{ijk}\left(b^{\hat{a}}_{~k}E^{\hat{a}}_{~j}
-\Omega^{\hat{a}}_{~p_{\bar{k}}x_k}b^{\hat{a}}_{~\bar{k}}eE_{j}\right)
\Bigr]n(\bm p) ,
\end{split}
\label{eq:polarization_3}
\eea
where $e>0$ is the electric charge. 
The first and second lines are the nonabelian correction to the Thouless pumping in the presence of the mixed-space Berry curvatures.
The nonabelian generalization of the Thouless pumping~\cite{PhysRevB.27.6083} contributes only to the spin pumping (See below).
The third line is the nonabelian chiral magnetic effect.
There are two contributions: One is the nonabelian generalization of the standard chiral magnetic effect, which survives even in the absence of the time-dependent perturbation.
The other is induced by external AC fields, and understood as the spectral flow in real space.
According to the axial anomaly equation, in the presence of nonzero $\bm E\cdot \bm  B$ or $\bm E^{\hat{a}}\cdot \bm  B^{\hat{a}}$, 
the electric charge is pumped in momentum space by the magnitude of $(\bm E\cdot \bm  B)\bm  b$ or $(\bm E^{\hat{a}}\cdot \bm  B^{\hat{a}})\bm  b$, 
which is the conventional spectral flow.
Since our kinetic theory is completely symmetric in phase space, in the presence of nonzero $\bm e\cdot \bm  b$ or $\bm e^{\hat{a}}\cdot \bm  b^{\hat{a}}$, 
the spectral flow arises in real space, and the electric charge is pumped in real space by the magnitude of $(\bm e\cdot \bm  b)\bm  B$ or
$(\bm e^{\hat{a}}\cdot \bm  b^{\hat{a}})\bm  B$.
The last line gives the nonabelian anomalous Hall effect, and the correction in the presence of the mixed-space Berry curvatures.
We remark here that there also exists the term, which comes from the anomalous velocity due to the Zeeman energy shift in the nonabelian magnetic field,
$j_{~i}^{\hat{0}}(t)=\int d^3p/(2\pi)^3\bm m^{\hat{a}}\cdot \bm B^{\hat{a}}(t)\p_{p_i}n(\bm p)$, 
where $\bm m^{\hat{a}}$ is the nonabelian magnetic moment. This is the nonabelian generalization of the gyrotropic magnetic effect~\cite{PhysRevLett.116.077201}.
However, the gyrotropic magnetic effect induces only the AC current, so that it vanishes in Eq.~\eqref{eq:polarization_3}.

Next the adiabatic spin pumping is given as
 \bea
 P_{~i}^{\hat{a}}(x_i)
  =&&2\lambda\int \frac{dtd^3p}{(2\pi)^3}j_{~i}^{\hat{a}}(t,\bm x,\bm p)
\nonumber \\
 =&&2\lambda\int \frac{dtd^3p}{(2\pi)^3}\Bigl[-\frac{1}{2}e^{\hat{a}}_{~i}
+\frac{1}{2}\left(\delta_{ij}\Omega^{\hat{a}}_{~x_kp_k}-\Omega^{\hat{a}}_{~p_ix_j}\right)\p_{p_j}\ve
\nonumber \\
-&&\frac{3}{40}\epsilon_{ikl}\epsilon_{j\bar{m}\bar{n}}\left(\Omega^{\hat{a}}_{~p_{\bar{m}}x_k}\Omega^{\hat{b}}_{~p_{\bar{n}}x_l}e^{\hat{b}}_{~j}
+\Omega^{\hat{b}}_{~p_{\bar{m}}x_k}\Omega^{\hat{a}}_{~p_{\bar{n}}x_l}e^{\hat{b}}_{~j}
+\Omega^{\hat{b}}_{~p_{\bar{m}}x_k}\Omega^{\hat{b}}_{~p_{\bar{n}}x_l}e^{\hat{a}}_{~j}\right)
\nonumber \\
+&&\frac{1}{2}eB_{i}\bm b^{\hat{a}}\cdot\nabla_{p}\ve
+\frac{3}{20}\left(B^{\hat{a}}_{~i} \bm e^{\hat{b}}\cdot \bm b^{\hat{b}}
+B^{\hat{b}}_{~i} \bm e^{\hat{a}}\cdot \bm b^{\hat{b}}
+B^{\hat{b}}_{~i} \bm e^{\hat{b}}\cdot \bm b^{\hat{a}}
\right)
\nonumber \\
+&&\frac{1}{2}\epsilon_{ijk}b^{\hat{a}}_{~k}eE_{j} 
-\frac{3}{20}\epsilon_{ijk}\left(\Omega^{\hat{a}}_{~p_{\bar{k}}x_k}b^{\hat{b}}_{~\bar{k}}E^{\hat{b}}_{~j}
+\Omega^{\hat{b}}_{~p_{\bar{k}}x_k}b^{\hat{a}}_{~\bar{k}}E^{\hat{b}}_{~j}
+\Omega^{\hat{b}}_{~p_{\bar{k}}x_k}b^{\hat{b}}_{~\bar{k}}E^{\hat{a}}_{~j} \right)
\Bigr]n(\bm p) ,
\label{eq:polarization_spin}
\eea
where $\lambda$ is the total angular momentum of the degenerate Bloch states~\cite{Murakami1348,PhysRevLett.93.156804}. 
The first and second lines are the nonabelian Thouless pumping and the correction in the presence of the mixed-space Berry curvatures.
The third line is the spin analogue of the chiral magnetic effect.
Only the first term persists in the absence of the external AC fields.
The last line gives the spin Hall effect, and the correction by the mixed-space Berry curvatures. 
In fact the first term reproduces the semiclassical result obtained in Refs.~\cite{Murakami1348,PhysRevLett.93.156804} in the absence of the time-dependent perturbation.

{\it Topological effective field theory.}  
We here show that  the abelian phase-space Chern-Simons theory derived in Refs.~\cite{PhysRevX.5.021018,Hayata:2016wgy} can be generalized to the nonabelian Berry curvatures.
We introduce the Chern-Simons Lagrangian density $\cL_{\rm CS}$ as the ($1+2d$) -form defined through 
\bea
d\cL_{\rm CS} &=& 
\frac{(-1)^\nu}{(2\pi)^d (d+1)!}c_{\hat{\nu}_1\ldots\hat{\nu}_{d+1}}\omega^{\hat{\nu}_1}\wedge\cdots\wedge\omega^{\hat{\nu}_{d+1}},
\label{eq:dCS_2d}
\eea
where 
$\omega^{\hat{\nu}}=\omega^{\hat{\nu}}_{~\alpha\beta}d\xi_\alpha\wedge d\xi_\beta/2=d\cA^{\hat{\nu}}+f^{\hat{\nu}\hat{\mu}\hat{\lambda}}\cA^{\hat{\mu}}\wedge\cA^{\hat{\lambda}}/2$
with $\cA^{\hat{\nu}}=\cA^{\hat{\nu}}_{~\alpha}d\xi_\alpha$.
We note that $c_{\hat{\nu}_1\ldots\hat{\nu}_{d+1}}$ is symmetric and invariant under the adjoint action of the Lie group. 
Then the Chern-Simons action is given as $S_{\rm CS}=\int_M \cL_{\rm CS}$, where $M$ is the entire manifold of phase space and time.
Although the Chern-Simons action itself has a complicated form for the nonabelian Berry curvatures in higher dimensions, the Chern-Simon current (i.e., the equation of motion) obtained from the variation of $S_{\rm CS}$ with respect to $\cA^{\hat{\nu}}_{~\mu}$ has a simple form (See e.g., Ref~\cite{Banados:1996yj}):
\bea
j_{~\mu}^{\hat{\nu}}&=&\frac{\p S_{\rm CS}}{\p {\cal A}_{~\mu}^{\hat{\nu}}}
\nonumber
\\
&=&\frac{(-1)^\nu}{(2\pi)^d2^{d} d!}c_{\hat{\nu}\hat{\nu}_1\ldots\hat{\nu}_d}\epsilon_{\mu \mu_1\ldots \mu_{2d}}\omega^{\hat{\nu}_1}_{~\mu_1\mu_2}\cdots  \omega^{\hat{\nu}_d}_{~\mu_{2d-1}\mu_{2d}} ,
\label{eq:CS_current_2d}
\eea
which recovers Eqs.~\eqref{eq:anomalouscurrent_2d} with $n(t,\xi_a) =1$ and~\eqref{eq:polarization_2d}. 
We find that the transports obtained from the kinetic theory can generally be expressed by the Chern-Simons current~\eqref{eq:CS_current_2d}. 
We here comment on the semiclassical approximation with respect to $q^{\hat{a}}$.
We treat $q^{\hat{a}}$ as the commutable variables, by considering the $\hbar\rightarrow0$ limit with $q^{\hat{a}}\cong \hbar t^{\hat{a}}$ fixed in $[\hbar t^{\hat{a}},\hbar t^{\hat{b}}]=i \hbar f^{\hat{a}\hat{b}\hat{c}} \hbar t^{\hat{c}}$, where $\hbar$ is the Planck constant.
Because of the classical limit (commutable $q^{\hat{a}}\cong \hbar t^{\hat{a}}$), 
the numerical factors of the Chern-Simons current~\eqref{eq:CS_current_2d} might not be correct except for the one in which $c_{\hat{\nu}_1\ldots\hat{\nu}_{d+1}}$ is the constant of motion, and corresponds to the Casimir invariant up to the normalization factor. 
From the comparison with the effective theory of the higher dimensional quantum Hall effect~\cite{PhysRevB.78.195424}, we infer that the full quantum results are recovered only by the modification of $c_{\hat{\nu}_1\ldots\hat{\nu}_{d+1}}$ in Eqs.~\eqref{eq:dCS_2d} and \eqref{eq:CS_current_2d} as $c_{\hat{\nu}_1\ldots\hat{\nu}_{d+1}}\rightarrow\sum_{\rm perm}\tr\left[t^{\hat{\nu}_1}\cdots t^{\hat{\nu}_{d+1}}\right]/(d+1)!$.
This discrepancy between the classical and quantum results would be resolved by the calculation based on the Wigner function formalism with accurately treating the star products of $q^{\hat{a}}$.

{\it Concluding remarks.} 
We have constructed the kinetic theory when all abelian and nonabelian phase-space Berry curvatures in $1+2d$ dimensions are nonzero,
by generalizing the one in the presence of the nonabelian gauge fields in real space to incorporate the interactions with the nonabelian gauge fields (Berry curvatures) in phase space.
Then we have calculated anomalous transports induced by the Berry curvatures. 
To show its utility, we have studied the anomalous transports induced by the SU($2$) Berry curvatures such as the nonabelian generalization of the adiabatic charge pumping, the chiral magnetic effect and the anomalous Hall effect, and the spin analogue of them. 
The transports given in Eqs.~\eqref{eq:polarization_3} and~\eqref{eq:polarization_spin} would arise in materials whose low-energy dynamics is described by e.g., the Luttinger model, which has the nonzero SU($2$) Berry curvatures.

We have also derived the effective theory to reproduce the transports in insulators calculated from the kinetic theory.
We have shown that such an  effective theory is the nonabelian generalization of the phase space Chern-Simons theory given in Ref.~\cite{PhysRevX.5.021018,Hayata:2016wgy}.
We have reveled the strong connection between the kinetic theory and the Chern-Simons theory, which is independent of spatial dimensions or the details of Berry curvatures.
Our kinetic theory provides the framework to study how the topological electromagnetic responses in insulators, which are usually analyzed by using the topological field theory, are modified in metals.

There are several future applications of our work.
One is a generalization to the nonequilibrium distribution.
We can calculate the anomalous transports in a steady state or in a non-adiabatic process by using the relaxation time approximation.
It is also interesting to consider the color-dependent distribution.
Another is to estimate the magnitude of the anomalous transports by using some model such as the Luttinger model. 
Although the advantage of our analysis is to predict the existence of the anomalous transports in a model independent way, 
it is also important to predict the magnitude of them for the implication to experiments.
Those analysis are left for future studies.

\begin{acknowledgements}

This work was supported by JSPS Grant-in-Aid for Scientific Research (No: JP16J02240).
This work was also partially supported by JSPS KAKENHI Grants Numbers 15H03652, 16K17716 and the RIKEN interdisciplinary Theoretical Science (iTHES) project.
\end{acknowledgements}


\bibliography{./cs}

\end{document}